\documentclass[12pt,draftclsnofoot,onecolumn]{IEEEtran}                                                        
\IEEEoverridecommandlockouts                              % This command is only
\def\BibTeX{{\rm B\kern-.05em{\sc i\kern-.025em b}\kern-.08em
		T\kern-.1667em\lower.7ex\hbox{E}\kern-.125emX}}
	
\usepackage[utf8]{inputenc}

\usepackage{epsfig,graphics,graphicx}
\usepackage[caption=false]{subfig}

\usepackage{amsmath}
\usepackage{amssymb}
\usepackage{float}
\usepackage{multirow}
\usepackage{bm}

\newcommand{\gauss}[2]{\mathcal{N}(#1,#2)} % Gaussian(1:mean,2:variance)
\newcommand{\cplxgauss}[2]{\mathcal{N}_{\mathbb{C}}(#1,#2)} % Complex-Gaussian(1:mean,2:variance)
\newcommand{\expected}[1]{\mathbb{E}\left\{#1\right\}} % Expected Value
\newcommand{\variance}[1]{\mathbb{V}\left\{#1\right\}} % Variance

\newcommand{\eye}[1]{\mathbf{I}_{#1}} % Identity matrix

\newcommand{\herm}{^\mathrm{\scriptscriptstyle H}} % hermitian
\newcommand{\transp}{^\mathrm{\scriptscriptstyle T}} % transpose
\newcommand{\cplxconj}{^\mathrm{*}} % complex conjugate
\newcommand{\norm}[1]{\left\lVert#1\right\rVert} % norm
\newcommand{\trace}[1]{\mathrm{tr}\left(#1\right)} % trace
\newcommand{\cplx}{\mathbb{C}} % complex Set

\newcommand{\tauc}{\tau_{\mathrm{c}}} % coherence length

\newcommand{\pilot}{{\mathrm{p}}} % pilot training phase
\newcommand{\taup}{\tau_{\pilot}} % pilot length
\newcommand{\rhop}{\rho^{\pilot}} % pilot transmitted power

\newcommand{\uplink}{{\mathrm{ul}}} % uplink phase
\newcommand{\rhoul}{\rho^{\uplink}} % uplink transmitted power

\title{Exponential Spatial Correlation with Large-Scale Fading Variations in Massive MIMO Channel Estimation}

\author{
	\IEEEauthorblockN{Victor Croisfelt Rodrigues\IEEEauthorrefmark{1}, Jos\'{e} Carlos Marinello Filho\IEEEauthorrefmark{1} and Taufik Abr\~{a}o\IEEEauthorrefmark{1}}\\
	\IEEEauthorblockA{\IEEEauthorrefmark{1} Department of Electrical Engineering (DEEL), State University of Londrina (UEL), Londrina, Brazil\\
	E-mail: victorcroisfelt@gmail.com, zecarlos.ee@gmail.com, taufik@uel.br}
}

\begin{document}
	
\maketitle
%\thispagestyle{empty}
%\pagestyle{empty}
%\vspace{-10mm}

\begin{abstract}
	To provide the vast exploitation of the large number of antennas on massive multiple-input–multiple-output (M-MIMO), it is crucial to know as accurately as possible the channel state information in the base station. This knowledge is canonically acquired through channel estimation procedures conducted after a pilot signaling phase, which adopts the widely accepted time-division duplex scheme. However, the quality of channel estimation is very impacted either by pilot contamination or by spatial correlation of the channels. There are several models that strive to match the spatial correlation in M-MIMO channels, the exponential correlation model being one of these. To observe how the channel estimation and pilot contamination are affected by this correlated fading model, this work proposes to investigate an M-MIMO scenario applying the standard minimum mean square error channel estimation approach over uniform linear arrays and uniform planar arrays (ULAs and UPAs, respectively) of antennas. Moreover, the elements of the array are considered to contribute unequally on the communication, owing to large-scale fading variations over the array. Thus, it was perceived that the spatially correlated channels generated by this combined model offer a reduction of pilot contamination, consequently the estimation quality is improved. The UPA acquired better results regarding pilot contamination since it has been demonstrated that this type of array generates stronger levels of spatial correlation than the ULA. In contrast to the favorable results in channel estimation, the channel hardening effect was impaired by the spatially correlated channels, where the UPA imposes the worst performance of this effect for the discussed model.
\end{abstract}


\section{Introduction}
In practice, the spatial directions of the channels that link a transmitter to receivers are defined by the arrangement of the propagation environment. Due to its nonregular disposition, it is congenitally possible to recognize that some spatial directions are more probable to carry strong signals than others \cite{Bjornson2017}. This preponderance is also imposed by the nonuniformity of the antenna radiation patterns, where, as a result of its irregularity, some spatial directions are more likely to receive extra power rather than others. Therefore, those phenomena give rise to a correlation degree between the gains and directions of the channels \cite{Bjornson2017}, in such a way, the power level of a channel is precisely related to its spatial orientation. This dependence is commonly nominated as spatial correlation, and its capacity to be obtained through a link will be referred to as spatiality. In general, this practical consequence can be exploited by massive multiple-input–multiple-output (M-MIMO) to obtain a better localization of the numerous user equipment (UE) served by a given system, when the UEs have adequately different spatial correlation behavior.

There are several approaches to model the spatiality of M-MIMO channels, being the exponential model one of the most common ways \cite{Loyka2001}. This model consists of an elementary design, which relies on a single parameter that controls the correlation level among the antennas over an array located at the base station (BS). Although simple, it has been shown to be useful to empirically represent the spatial correlation in uniform linear arrays (ULA) \cite{Loyka2001}. Notwithstanding, it can also be used to approximately model the spatial correlation of uniform planar array (UPA) arrangements by virtue of the Kronecker product \cite{Ying2014}, providing a way to separate the horizontal and vertical dimensions of the UPA with co-polarized antennas. Recently, \cite{Gao2015} and \cite{Gao2015a} showed that the power received by each antenna of an array varies arbitrarily, emanating a nonequal contribution of each antenna to a given communication link. This phenomenon also contributes to spatial correlation and its concern represents a more realistic condition from the deployment point of view.

The exploitation of the elevation dimension upon 2-dimensional (2D) antenna arrays has proliferated as a notable enhancement to handle the high data rates and reliability expected for the fifth generation (5G) of wireless communication. This is possible through the implementation of multiantenna techniques, as elevation beamforming and full-dimension MIMO. However, as a counter point, the 2D antenna arrays demand a more troublesome 3-dimensional (3D) characterization of the channel models. The most common approaches of 2D antenna arrangements are through circular or planar arrays \cite{Bjornson2017}. In particular, the UPA has received a great appeal because of its compact assemble and the possibility to build it on the facade of buildings. In contrast with the ULA that only supports signal reception on the azimuth direction, UPA can provide control of the communication on both azimuth and elevation dimensions. Thus, it can be inferred that the investigation of UPA under spatial correlation and how M-MIMO reacts to it are important questions that are needed to be answered. The performance comparison between ULA and UPA is a pertinent discussion as well.

The better usage of the large antenna arrays is reliant on the estimation of the channel vectors that link the UEs to their respective BSs. The standard least-squares (LS) and minimum mean square error (MMSE) estimators have been widely deployed to acquire the channel state information in the BSs, under the realization of a pilot training phase commonly occurring in the time-division duplex (TDD) mode. Generally, the pilot signals used in the estimation process are reused across the UEs, due to the increasing number of UEs and constraints related to the characteristics of the wireless channels. This reuse contributes to the derivation of an undesirable interference in the channel estimation phase, which is known as pilot contamination \cite{Marzetta2016}. In view of the fact that the spatial correlation affects the covariance matrices of the channel responses, the channel estimation and the effect of pilot contamination are duly affected by the spatially correlated channels. It is therefore important to obtain some insights on how the spatial correlation can influence the estimation process when considering the ULA and UPA arrangements. 

By joining the motivations given above, it is evident to perceive that a great interest exists in understanding the spatial correlation effect on M-MIMO since it is a practical operation circumstance. This concern has been established as the subject of several works through the years. As an example, \cite{Bjornson2017} examined the spatial correlation effect onto the channel estimation considering a one-ring model, which commonly gives support to rank-deficient channel covariance matrices. Considering the exponential correlation model, the estimation quality of the MMSE channel estimation was analyzed for M-MIMO when applying a ULA in \cite{Albdran2016}. However, it does not consider UPA nor even large-scale fading variations over the array; besides, it did not give any insight about how the pilot contamination is affected by the spatial correlation. In \cite{Bjornson2018}, it was demonstrated that the pilot contamination vanishes when the number of antennas goes to infinity under the consideration of an exponential correlation model with large-scale fading variations over the array. Nevertheless, it did not demonstrate some statistical characteristics of this model and only a ULA arrangement was considered. The derivation of superior and inferior analytical bounds for the eigenvalues generated by the exponential correlation model is treated in \cite{Choi2014} and \cite{Lim2017}. However, the authors did not discuss the influence of this model over M-MIMO scenarios. A more robust analytical study behind the eigenstructure of the exponential correlation matrix is carried out in \cite{Mallik2018}. Some insights on the performance of a wireless system are offered in such work, but it did not consider massive antenna arrays.

This work aspires to remedy the investigation of how the spatial correlation affects the estimation of the channel responses assuming a more general model of the spatial correlation effect in covariance matrices. This design is based on the exponential correlation model combined with the large-scale fading variations over the array, which compound is seldom investigated to the best of our knowledge. At the first moment, this model is properly characterized through an analysis of the distribution of the eigenvalues of the covariance matrices generated for ULA and UPA. Then, we strive to seek for more consistent results on the spatiality impact over UPA, adopting the channel hardening and favorable propagation effects as metrics. The quality of MMSE channel estimates\footnote{The use of the MMSE estimator makes neat the analysis of spatial correlation, since it does not carry out the imperfections obtained by the LS estimator.} is investigated under spatially correlated channels considering ULA and UPA, with these results being compared to the canonical uncorrelated full rank i.i.d. Rayleigh fading channel. The evaluation of the obtained results permits a good comparison between the ULA and UPA regarding the uncorrelated and correlated scenarios, uncovering the benefits and drawbacks of each arrangement that are important for M-MIMO network design.

The remainder of this paper is organized as follows. Section \ref{sec:systemmodel} introduces the M-MIMO system model, where the focus is to derive the expressions for pilot training and MMSE channel estimation. The exponential correlation model with large-scale fading variations over the array is discussed in Section \ref{sec:exponential}, as well as the influence of this compound model over channel hardening and favorable propagation. Numerical results demonstrating the impact of spatial correlation over the channel estimation and pilot contamination are developed in Section \ref{sec:spatialcorr-chnest}. The main conclusions are summarized in Section \ref{sec:conc}.

\textit{Notations}: The superscripts $(\cdot)\transp$, $(\cdot)\herm$ and $(\cdot)\cplxconj$ denote the transpose, Hermitian transpose, and complex conjugate, respectively. The $N \times N$ identity matrix is indicated by $\eye{N}$. The $\cplxgauss{\cdot}{\cdot}$ stands for the circularly symmetric complex Gaussian distribution. The statistical operators $\expected{X}$ and $\variance{X}$ represent the expected value of a random variable $X$ and its variance, subsequently. The operator $\otimes$ represents the Kronecker product, whereas $\trace{\cdot}$ stands for the trace of a matrix.

\section{System Model}\label{sec:systemmodel}
The pilot training phase of a canonical M-MIMO system is presented along this section, where the pilot signaling is realized through the use of a TDD scheme \cite{Bjornson2017,Marzetta2016}. The system consists of $L$ cells working synchronously\footnote{The synchronous operation provides a worst-case scenario from the point of view of pilot contamination, due to the fact that all cells are in pilot training at the same time.} and sharing the same time-frequency resources. Each BS is equipped with $M$ antennas and it serves $K$ single-antenna UEs. In our context, the pilot sequences attributed to the UEs are being reused across the cells, as a result of the difficulty to generate a large number of orthogonal pilot sequences inside a short coherence interval $\tauc$. However, the UEs within a same cell have mutually orthogonal pilot sequences with the goal of suppressing the intracell interference.

Let $\displaystyle \mathbf{g}_{jlk} \in \cplx^{M}$ denotes the channel from UE $k$ within cell $l$ to BS $j$. Assuming a correlated Rayleigh fading model, the channel can be written as $\mathbf{g}_{jlk} \sim \cplxgauss{\mathbf{0}}{\mathbf{R}_{jlk}}$; where the covariance matrix, $\mathbf{R}_{jlk} \in \mathbb{C}^{M \times M}$, embodies effects such as pathloss, shadowing, and spatial channel correlation, which correspond to the large-scale propagation phenomena, whereas the complex Gaussian distribution stands for the small-scale fading. One should note that the eigenstructure of $\mathbf{R}_{jlk}$ expresses the spatial channel correlation, whose phenomenology is mathematically indicated by the nonzero off-diagonal elements of the covariance matrix. Here, the diagonal entries of $\mathbf{R}_{jlk}$ are also considered nonidentical, stressing the possibility of large-scale fading variations over the array \cite{Bjornson2018}. Henceforth, it will be assumed that the covariance matrices are perfectly known for anyone who needs to know them.\footnote{The second moment estimation that concerns with the imperfect knowledge of the covariance matrices is studied by many works (see \cite{Bjornson2017b}, for example), but it is beyond the scope of this work.} 

\subsection{Uplink Pilots}
During the pilot training, all UEs transmit their $\taup$-length pilots to their corresponding BSs, where both link ends have a prior knowledge of these pilots. The typical assumption of $\taup = K$ will be considered herein, whereby the UE $i$ of each cell reuses the same pilot, while the UEs assigned with different pilots are mutually orthogonal. That being the case, the
received pilot signal in BS $j$, $\mathbf{Y}^{\mathrm{p}}_{j} \in \mathbb{C}^{M \times \taup}$, can be expressed as
\begin{equation}
\mathbf{Y}^{\mathrm{p}}_{j} = \sqrt{\rhop} \sum_{l = 1}^{L} \sum_{k = 1}^{K} \mathbf{g}_{jlk} \boldsymbol{\phi}_{k}\transp + \mathbf{N}^{\mathrm{p}}_{j},
\label{eq:pltrxsignal}
\end{equation}
where $\rhop$ is the normalized pilot transmit power with normalization relative to the magnitude of the receiver noise, and $\boldsymbol{\phi}_{k} \in \cplx^{\taup}$ is the pilot assigned to the $k$th UE within cell $l$. The pilot is considered to be normalized in such a way that it does not intervene on the transmitted power. Hence, it is highlighted that the pilot has unit norm: $\textstyle \norm{\boldsymbol{\phi}_{k}}^2 = \boldsymbol{\phi}_{k}\herm \boldsymbol{\phi}_{k} = 1$. $\mathbf{N}^{\mathrm{p}}_{l} \in \cplx^{M \times \tau_{p}}$ is a matrix with the normalized receiver noise with each element obeying an i.i.d. $\cplxgauss{0}{1}$.

Preceding the estimation process, perceive that \eqref{eq:pltrxsignal} can be written in such a manner that culminates the drop of its dependence on pilots, resulting in a signal only dependent of the channels and noise. Therefore, augmenting that the BS $j$ wants to estimate the channel of UE $i$ attached to BS $l$, the following computation is performed: 
\begin{equation}
	\mathbf{y}^{\mathrm{p}}_{jli} = \mathbf{Y}^{\mathrm{p}}_{j} \boldsymbol{\phi}_{i}\cplxconj = \sqrt{\rhop} \sum_{l' = 1}^{L} \mathbf{g}_{jl'i} + \bar{\mathbf{n}}^{\mathrm{p}}_{ji},
\label{eq:plt-uncorr}
\end{equation}
where $\mathbf{y}^{\mathrm{p}}_{jli} \in \cplx^{M \times 1}$ and $\bar{\mathbf{n}}^{\mathrm{p}}_{ji} \in \cplx^{M \times 1}$ is distributed as $\cplxgauss{\mathbf{0}_{M\times{1}}}{\eye{M}}$, since the realized operation is linear and unitary. Then, a prominent remark can be made given that $\rhop = \taup \rhoul$, where $\rhoul$ is the normalized uplink transmit power. It is therefore possible to note that a wider pilot signaling phase improves the estimation quality, which will be more clearly evident in the following section.

\subsection{MMSE Channel Estimation}
The Bayesian's MMSE estimator usually obtains optimal estimation performance, thanks to the exploitation of prior statistical knowledge of the physical event \cite{Kay1993}. Straightforwardly, the channel estimate obtained in BS $j$ for the $i$th UE within cell $l$ is expressed as \cite{Bjornson2018,Kay1993}
\begin{equation}
	\hat{\bf g}_{jli} = \sqrt{\rhop} \mathbf{R}_{jli} \mathbf{Q}^{-1}_{ji} \mathbf{y}^{\mathrm{p}}_{jli} \sim \cplxgauss{\mathbf{0}}{\boldsymbol{\Psi}_{jli}},
	\label{eq:mmse:estimate}
\end{equation}
where $\textstyle \mathbf{Q}_{ji}=\rhop \sum_{l'=1}^{L} \mathbf{R}_{jl'i} + \eye{M}$ denotes the variance of the observation given in \eqref{eq:plt-uncorr}, and $\boldsymbol{\Psi}_{jli} = \rhop \mathbf{R}_{jli} \mathbf{Q}^{-1}_{ji} \mathbf{R}_{jli}$ is the covariance matrix of the estimate. The estimation error is specified as $\tilde{\mathbf{g}}_{jli}=\mathbf{g}_{jli} - \hat{\mathbf{g}}_{jli}$, which is a random quantity distributed as $\cplxgauss{\mathbf{0}}{\mathbf{C}_{jli}}$ with $\mathbf{C}_{jli} = \mathbf{R}_{jli} - \boldsymbol{\Psi}_{jli}$. Relying on the orthogonality principle of the MMSE estimator, it is said that the estimate and the estimation error are independent quantities. It should be noted that the UEs assigned with the same pilot through the cells have their estimates correlated. In fact, the correlation between a peer of UEs can be measured via the antenna-averaged correlation coefficient. This metric suitably determines the association level of two interfering channels per antenna; one channel is referent to the $i$th interfering UE within the $l$th cell and another to the desired $k$th UE attached to BS $j$. Thus, we have \cite{Bjornson2017}
\begin{equation}
	{c_{jk,li}}=\dfrac{\mathbb{E}\{\hat{\mathbf{g}}^{\mathrm{\scriptscriptstyle H}}_{jli}\hat{\mathbf{g}}_{jjk}\}}{\sqrt{\mathbb{E}\{\lVert{\hat{\mathbf{g}}_{jli}}\rVert^2\}\mathbb{E}\{\lVert{\hat{\mathbf{g}}_{jjk}}\rVert^2\}}}=\dfrac{\mathrm{tr}({\mathbf{R}_{jli}\mathbf{R}_{jjk}\boldsymbol{\Psi}_{jli}})}{\mathrm{tr}({\mathbf{R}_{jli}\mathbf{R}_{jli} \boldsymbol{\Psi}_{jli}})\mathrm{tr}({\mathbf{R}_{jlk}\mathbf{R}_{jjk}\boldsymbol{\Psi}_{jli}})},
	\label{eq:mmse:ant-avgcorrcoeff}
\end{equation}
where the closer $c_{jk,li}$ is from $0$, lower is the pilot contamination provided by the interfering UE. Otherwise, if $c_{jk,li}$ is close to $1$, the interfering UE dramatically impacts the channel estimation of UE $k$.

Intending to evaluate the estimation quality, it is interesting to present the normalized mean squared error (NMSE) expressed as \cite{Bjornson2017}
\begin{equation}
	\mathrm{NMSE}_{jli} = \dfrac{\mathbb{E}\{\lVert{\tilde{\mathbf{g}}_{jli}}\rVert^2\}}{\mathbb{E}\{\lVert{\mathbf{{g}}_{jli}}\rVert^2\}} = \dfrac{\mathrm{tr}({\mathbf{C}}_{jli})}{\mathrm{tr}({\mathbf{R}}_{jli})}.
	\label{eq:mmse:nmse}	
\end{equation}
This definition specifies the relative error of channel estimation, which varies between $0$ and $1$ per antenna. The former value represents the perfect estimation, whereas the latter indicates a worst-case scenario that is achieved when the mean value of the channel is used as an estimate \cite{Bjornson2017}.

\section{Exponential Correlation Model with Large-Scale Fading Variations over the Array}\label{sec:exponential}
The exponential correlation model is a simple single-parameter representation of spatially correlated channels that stems from the antenna array arrangement. It is based on a correlation coefficient given as $r \in [0,1]$, which indicates the level of how much a peer of antennas over the array is spatially correlated \cite{Loyka2001}. The model spans a smaller correlation level for distant antennas by taking into consideration the value of $r$, whereas the close antennas have a higher correlation. These considerations are quite physically intuitive and are strongly related to the spacing between antennas, being validated for some environments through empirical measurements of MIMO systems, as reported in \cite{Loyka2001}. 

Motivated by measurements campaigns \cite{Gao2015,Gao2015a}, here, the well-known exponential correlation model is considered with the addition of independent log-normal large-scale fading variations over the array. These fluctuations can be interpreted as if each antenna is affected by a different realization of shadow-fading, due to the arbitrary disposition of the scatters in the propagation medium. In view of this, the ($m,n$)th element of the channel covariance matrix can be written as \cite{Loyka2001,Bjornson2018}
\begin{equation}
\left[\mathbf{R}\right]_{m,n} = \beta r^{n-m} e^{i(n-m)\theta} 10^{(f_{m} + f_{n} / 20)}
\label{eq:exp:ula-exponential}
\end{equation}
where {$m,n\in\{1,\dots,M\}$}, $\beta$ is the average large-scale fading coefficient (pathloss), $r\in[0,1]$ is the correlation factor, $\theta \in [-\pi,+\pi)$ is the angle-of-arrival (AoA) and $f_1, \dots, f_{M} \sim \gauss{0}{\sigma^2}$ forms the random fluctuations of the medium-scale fading (shadowing) with standard deviation $\sigma$. Observe that the subscripts of $\mathbf{R}$ were omitted to facilitate the description. 

In fact, it can be noted that the resultant model is a double-parameter model described by $r$ and $\sigma$. The above model fits well the description of a ULA, where it will be considered an horizontal ULA with $\theta \in [-\pi,+\pi)$ representing the AoA on the azimuth direction. Moreover, notice that a strong spatial correlation is identified by large variations of the eigenvalues of $\mathbf{R}$.

Despite the exponential correlation model expressed in \eqref{eq:exp:ula-exponential} is immediately valid to ULA, it is not well founded to UPA arrangements. Fortunately, the authors in \cite{Ying2014} demonstrated that a UPA channel model can be approximated from the ULA counterpart by using a Kronecker model. This result was supported by the use of a ray-based channel model. Then, the 3D channel modeling for UPA can be seen as two dissociated 2D channels on the azimuth and elevation directions. In other words, the spatial correlation matrix for UPA can be approached as two ULA, one disposed horizontally and the other vertically, making it possible to write the following channel covariance matrix for UPA \cite{Ying2014}:
\begin{equation}
\mathbf{R} \approx \mathbf{R}_{h} \otimes \mathbf{R}_{v} 
\label{eq:exp:upa-exponential}
\end{equation}
where it should be assumed as the usage of co-polarized antennas. $\mathbf{R}_{h} \in \cplx^{M_{h} \times M_{h}}$ and $\mathbf{R}_{v} \in \cplx^{M_{v} \times M_{v}}$ are, respectively, the horizontal and vertical correlation matrices related to their respective ULA components. Notice that the values of $M_{h}$ and $M_{v}$ must be put in such a way that $M = M_{h}M_{v}$. One should emphasize that the vertical AoA is denoted by $\varphi \in [-\pi/2,+\pi/2)$, which indicates the support of UPA on the elevation dimension. As before, the horizontal AoA is $\theta \in [-\pi,+\pi)$ corresponding to the azimuth orientation. Additionally, a useful property of the Kronecker product that will be used on the further discussions is stated in the following definition.

\textbf{Definition 1.} Since $\mathbf{R}_{h} \in \cplx^{M_{h} \times M_{h}}$ and $\mathbf{R}_{v} \in \cplx^{M_{v} \times M_{v}}$ are square matrices with eigenvalues denoted as $\lambda_{\mathbf{R}_{h},i}, i = 1,2,\dots,M_{h}$ and $\lambda_{\mathbf{R}_{v},k}, \ k = 1,2,\dots,M_{v}$, respectively, then it is demonstrable that $\mathbf{R}_{h} \otimes \mathbf{R}_{v}$ implies by its multiplicity property that $\lambda_{\mathbf{R}_{h},i}\lambda_{\mathbf{R}_{v},k}, \ \forall i,k$, i.e., the Kronecker product between two matrices culminates in a matrix that has, as its eigenvalues, the multiplication of the eigenvalues of the two primary matrices.

\subsection{Illustrative Results for ULA}
\begin{figure}[htp]	
	\centering
	\subfloat[Different values of antenna correlation $r$ and $\sigma = 0$ dB.]{%
		\includegraphics[trim={6mm 2mm 6mm 5mm}, clip, width=.75\textwidth]{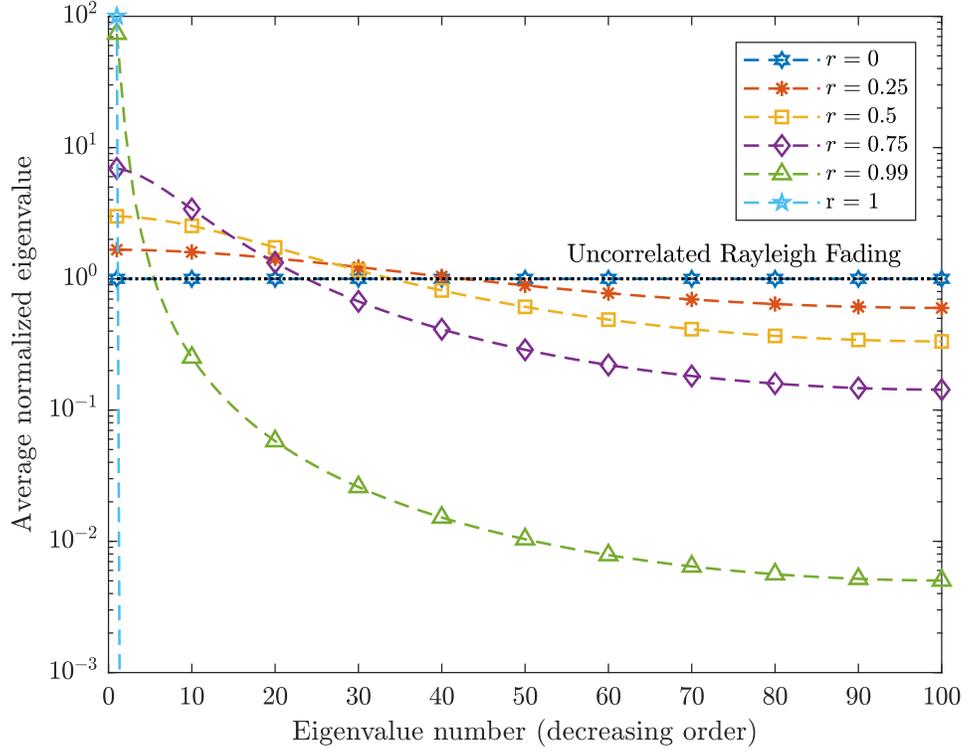}%
		\label{fig:exp:ula:eigenvaluedist:r}
	}
	
	\subfloat[Different values of shadowing standard deviation $\sigma$ and $r=0$.]{%
		\includegraphics[trim={6mm 2mm 6mm 5mm}, clip, width=.75\textwidth]{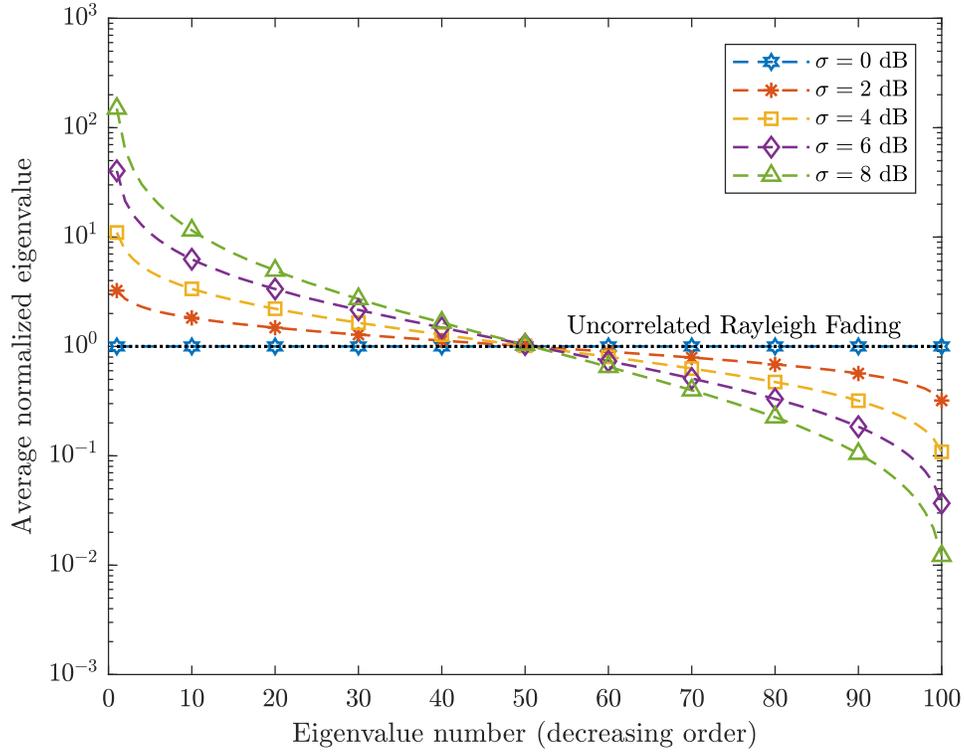}%
		\label{fig:exp:ula:eigenvaluedist:std}
	}
	
	\caption{Normalized distribution of the eigenvalues ($\beta=1$) as a function of the antenna indexes sorted in a decreasing order for an $M = 100$ ULA, (a) several values of $r$ and fixing $\sigma$ and (b) several values of $\sigma$ and fixing $r$. The uncorrelated Rayleigh fading was plotted as a reference.}
	\label{fig:exp:ula:eigenvaluedist}
\end{figure}

To gain further insights into the covariance matrices generated by the presented model, it is now aimed to investigate the normalized \footnote{The term "normalized" stands for the usage of $\beta = 1$.} distribution of their eigenvalues. For this purpose, it is considered a scenario comprised of a ULA with $M = 100$ antennas. Once the discussed model is a double-parameter model, the analysis will be split into two extreme cases: one with $r$ varying when $\sigma = 0$ dB and another with the previous statement opposite. These extreme cases are depicted in Fig. \ref{fig:exp:ula:eigenvaluedist}. As expected, it can be seen from Fig. \ref{fig:exp:ula:eigenvaluedist:r} that the increase in $r$ provides stronger spatial correlation since $r$ indicates the interdependence between the antennas of the array. In addition, when the standard deviation of the random fluctuations over the array increases, the spatial correlation strength appears to grow, as indicated in Fig. \ref{fig:exp:ula:eigenvaluedist:std}. The explanation of this last result is in the capacity of some arbitrary directions to have more power than others, in view of the shadow-fading be differently experienced by each antenna. In other words, when the shadowing variability, $\sigma$, increases, more power is allocated quasi-consistently to some antennas, as can be seen by the almost symmetric curve onto the vertical axis positioned at the fiftieth antenna index in Fig. \ref{fig:exp:ula:eigenvaluedist:std}. It is worth saying that the shapes of the curves are a straight consequence of the log-normal distribution. Finally, one can stress that the compound of exponential correlation model with large-scale fading variations over the array for ULA does not provide rank-deficient matrices in most of the practical cases, as the one-ring model does for ULA \cite{Bjornson2017,Bjornson2018}. It is important to indicate that this property is considered a special case in practice and our concern here is for a more general framework. However, rank-deficient matrices are provided for the unexpected cases of $r=1$ or very high values of $\sigma$.

\subsection{Illustrative Results for UPA}
\begin{figure}[htp]
	\centering
	\subfloat[Different values of antenna correlation $r$ and $\sigma = 0$ dB.]{%
		\includegraphics[trim={6mm 2mm 6mm 5mm}, clip, width=.75\textwidth]{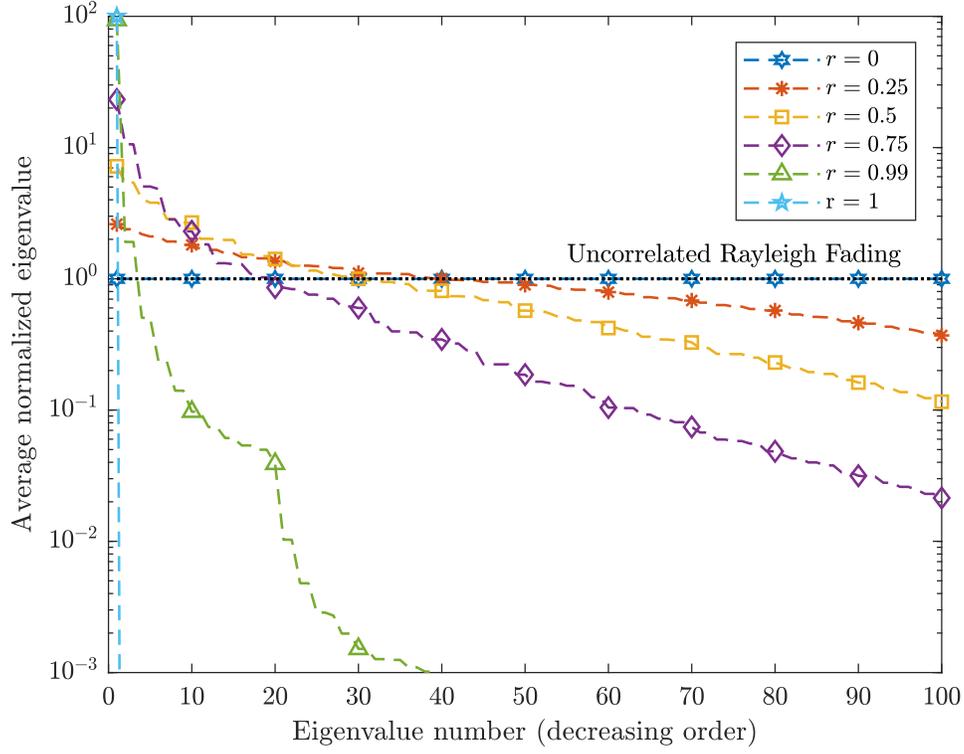}%
		\label{fig:exp:upa:eigenvaluedist:r}
	}
	
	\subfloat[Different values of shadowing standard deviation $\sigma$ and  $r=0$.]{%
		\includegraphics[trim={6mm 2mm 6mm 5mm}, clip, width=.75\textwidth]{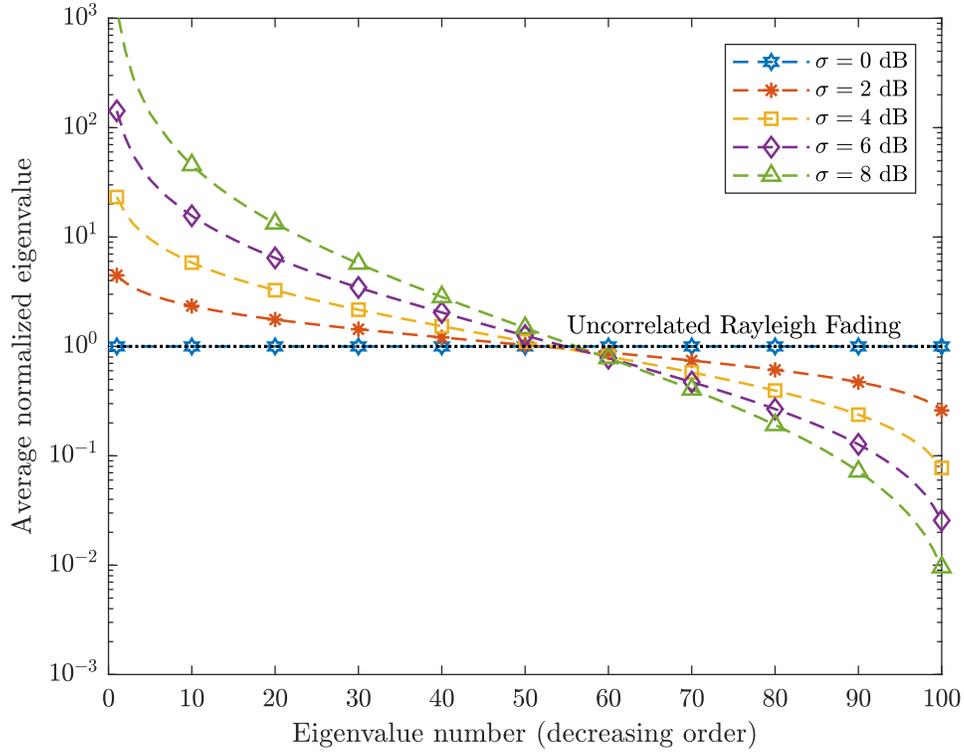}%
		\label{fig:exp:upa:eigenvaluedist:std}
	}
	
	\caption{Normalized {distribution of the} eigenvalues ($\beta=1$) as a function of the antenna indexes sorted in a decreasing order for {an} $M$ = 100 square UPA, (a) several values of $r$ and fixing $\sigma$ and (b) several values of $\sigma$ and fixing $r$. The uncorrelated Rayleigh fading was plotted as a reference.}
	\label{fig:exp:upa:eigenvaluedist}
\end{figure}

As before, the normalized distributions of the eigenvalues embraced by the covariance matrices were conceived assuming a square UPA with $M_{h}=M_{v}=10$ and, hence, $M = 100$. These results are shown in Fig. \ref{fig:exp:upa:eigenvaluedist} for the extreme cases of $\sigma=0$ dB and $r=0$. As can be seen from Fig. \ref{fig:exp:upa:eigenvaluedist:r}, the curves have similar shapes to the ULA case, but they are more distorted, presenting higher eigenvalues than the previous scenario. Furthermore, it is possible to note a greater amount of power allocated to a few number of antennas and the fact that smaller is the amount assigned to the other antennas. These effects are directly explained through Definition 1, whereby the power received (eigenvalues) on the azimuth direction is being associated with the acquired on the elevation orientation. Since the square arrangement has a more confined space than the linear counterpart and prompted by the fact that $r$ plays a more strong effect in confined spaces, it is then possible to conclude that the UPA generates stronger spatial correlation levels than the ULA. 

Fig. \ref{fig:exp:upa:eigenvaluedist:std} was conceived for the case that $r = 0$, i.e., the effect of $\sigma$ is now being evaluated. In contrast to the ULA scenario, it can be observed that the curves are not more quasi-symmetric. Recall that as shown from Fig. \ref{fig:exp:ula:eigenvaluedist:std}, the quasi-symmetry stands for the fact that the quantity of antennas with high fraction of power is almost equal to the number of antennas {assigned to the small part of power}. This different finding can be mathematically explained by using Definition 1 {and} resorting to {the} multiplication of two log-normal random variables. For the sake of demonstration, consider two independent log-normal random variables $X\sim\exp(\gauss{0}{\sigma_{x}^2})$ and $Y\sim\exp(\gauss{0}{\sigma_{y}^2})$. The Kronecker product comprises of {the multiplications of the main diagonals}, where these diagonals {are composed of} values distributed like $X$ and $Y$. It is easy to demonstrate that the multiplication of $Z = XY$ yields {in} $Z\sim \exp(\gauss{0}{\sigma_{x}^2+\sigma_{y}^2})$. Hence, observe that the large-scale fading variations over the array for UPA is obeying an $\exp(\gauss{0}{2\sigma^2})$ distribution. 
Based on this latter result, it is possible to sustain the idea that some antennas of the UPA are receiving a similar shadow-fading effect through some directions, that is, from the azimuth and elevation orientations. Consequently, the shadowing constructively sums up across the antenna elements, leading to a more quantity of antennas containing a great amount of power. This hypothesis is supported by the right shift of the curves presented in Fig. \ref{fig:exp:ula:eigenvaluedist:std} with respect to the given in  Fig. \ref{fig:exp:upa:eigenvaluedist:std}, along with the fact that more antennas are assigned with pertinent power quantities. In summary, the shadow-fading for UPA has a greater variance than the observed for ULA, due to the antenna arrangement of the former be more confined.

Although it appears that rank-deficient matrices are more likely to occur in the UPA scenario than ULA, the variations between the higher and smaller eigenvalues on UPA are not so great to sustain a quite relevant difference in spatiality that could cause the special case of rank deficiency. It is therefore possible to say that the rank-deficient matrices under the consideration of UPA are obtained in conditions similar to those aforementioned for the ULA.

\subsection{Channel Hardening and Favorable Propagation}
This section briefly calls attention to the effects of the spatiality generated by the exponential model with large-scale fading variations over channel hardening and favorable propagation effects. It is known that multiantenna systems provide these two desirable attributes, which support several benefits to the performance of {M-MIMO} systems \cite{Bjornson2017}. In fact, channel hardening and favorable propagation are very related to the characteristics of the propagation environment, such as the spatial correlation. Furthermore, one should stay clear that these properties can be attained jointly or disjointly, even being not assured for some propagation environments.

Accordingly to \cite{Bjornson2017}, the channel hardening effect is reduced with the spatial correlation increase. Thus, one can presume that the channel hardening reduces as $r$ and $\sigma$ increase. This attribute consequently seems to be very unsatisfactorily sustained for UPA arrangements since UPA generates stronger spatial correlation {in comparison to the ULA}. On the other hand, the favorable propagation is prone to increase with spatial correlation, if the UEs have distinct spatial characteristics \cite{Bjornson2017}. Recall that favorable propagation underlies the environment capacity in aid sufficiently distinct directions of the channels from UEs to BS. Therefore, one can infer the opposite regarding the channel hardening for favorable propagation, in view of the fact that the spatial correlation generally assists the channel differentiation of the UEs. This last effect is indirectly investigated in the next section, when the impact of the spatiality over channel estimation and pilot contamination are analyzed. That being said, in this section, only the channel hardening is numerically evaluated to give some insights about its behavior under the considered spatial correlation model.

\begin{figure}[htbp]
	\centering
	\subfloat[Variance {\it vs}  $M$ for some values of $r$ and $\sigma=0$ dB.]{
		\includegraphics[trim={6mm 2mm 6mm 5mm}, clip, width=.75\textwidth]{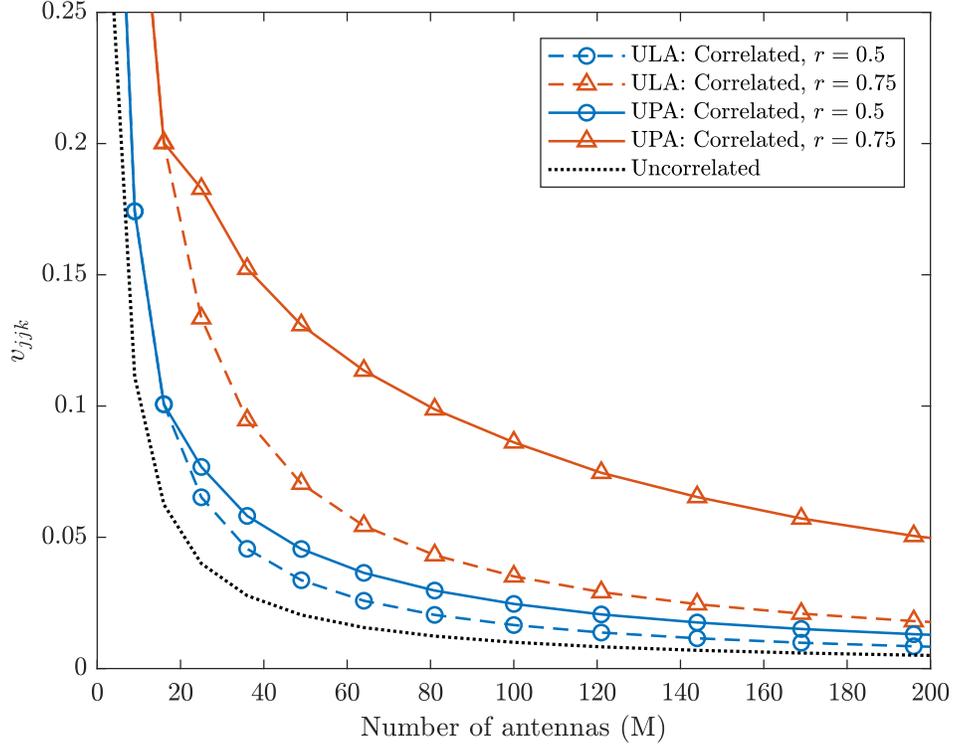} 
		\label{fig:chnhard:r}
	}
	
	\subfloat[Variance {\it vs} $M$ for some values of $\sigma$ and $r=0$.]{
		\includegraphics[trim={6mm 2mm 6mm 5mm}, clip, width=.75\textwidth]{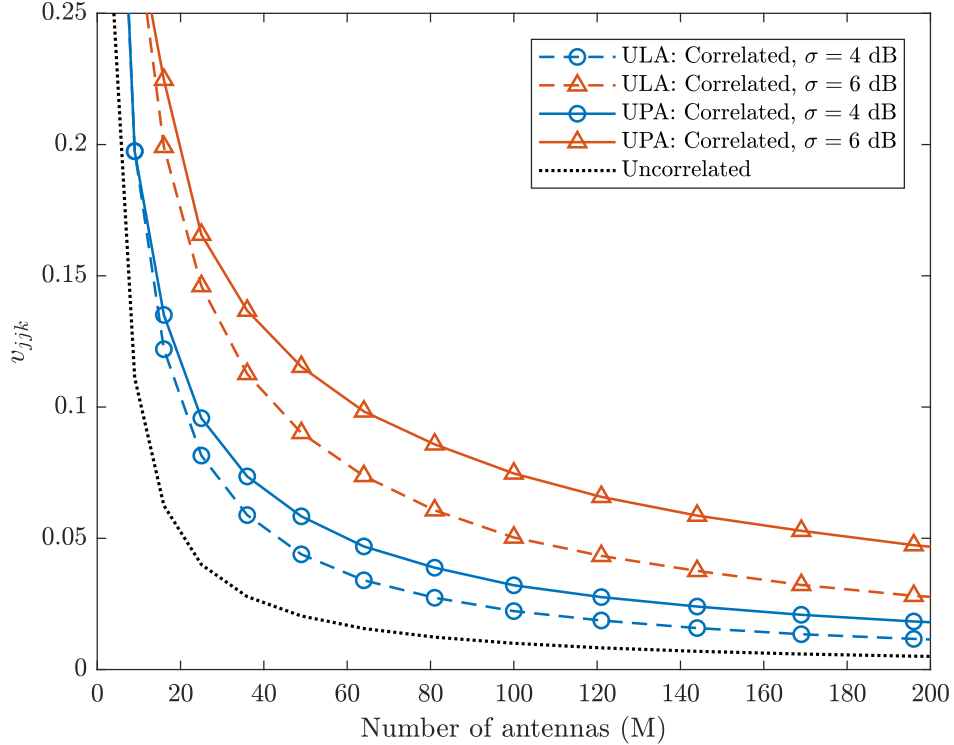}
		\label{fig:chnhard:std}
	}
	\caption{Variance $v_{jjk}$, eq. \eqref{eq:variance:chnhardening}, as a function of $M$; considering (a) different values of $r$ when $\sigma=0$ dB and (b) different values of $\sigma$ when $r=0$. The curves were obtained supposing that the desired UE $k$ is positioned at $(\theta,\varphi)=(30^{\mathrm{o}},30^{\mathrm{o}})$. For the case that $\sigma$ varies, each point was taken in average for $1000$ realization of an ergodic process.}  
	\label{fig:chnhard}
\end{figure}

The spatial diversity consists of the fact that each receiver antenna is experiencing an independent fading realization. In this case, the small-scale fading can be progressively tackled over a given communication link because the concatenation of the multiple received signals turns out to behave almost as a deterministic quantity due to its variability decrease. This last result is then nominated as channel hardening, which is expected to be more effectively obtained when there is a large number of antennas. Several improvements are associated with this phenomenon, such as the mitigation of small-scale fading and the tightness of spectral efficiency equations for M-MIMO. The asymptotic channel hardening through a given channel linking UE $k$ within cell $j$ to BS $j$ can be defined as \cite{Bjornson2017} 
\begin{equation}
	\dfrac{\lVert{\mathbf{g}_{jjk}}\rVert^2}{\mathbb{E}\{\lVert{\mathbf{g}_{jjk}}\rVert^2\}} \xrightarrow[M \rightarrow \infty]{a.s.} 1.
	\label{eq:chnhardening}
\end{equation}
This expression indicates that the channel gain, given by the numerator, shows a tendency to be close of its mean value, the denominator, when the number of antennas goes large. In addition, notice that this relation is kept true for any fading channel $\mathbf{g}_{jjk}$ \cite{Bjornson2017}. In this way, one can conclude that the channel gains are prone to be more confined in values near its mean, which means that the channel variability vanishes for $M \rightarrow \infty$. Alternatively to \eqref{eq:chnhardening}, one can write an expression to quantify how close a given channel is near to the channel hardening condition, regardless of the asymptotic analysis. Thereby, the following variance measure is able to capture this effect \cite{Bjornson2017}:
\begin{equation}
	v_{jjk} = \mathbb{V}\left\{\dfrac{\lVert{\mathbf{g}_{jjk}}\rVert^2}{\mathbb{E}\{\lVert{\mathbf{g}_{jjk}}\rVert^2\}}\right\} = \dfrac{\mathbb{V}\{\lVert{\mathbf{g}_{jjk}}\rVert^2\}}{\left(\mathbb{E}\{\lVert{\mathbf{g}_{jjk}}\rVert^2\}\right)^2} = \dfrac{\mathrm{tr}\left({\mathbf{R}_{jjk}^2}\right)}{\left( \mathrm{tr}({\mathbf{R}_{jjk})}\right)^2},
	\label{eq:variance:chnhardening}
\end{equation}
where observe that the more the variance approaches zero, the more harden is the channel. The numerator above gives the sum of the squared eigenvalues, whereas the denominator represents the squared sum of the eigenvalues \cite{Bjornson2018}. Therefore, it is conceivable to infer that the spatial correlation affects channel hardening since the eigenvalues are impacted, as could be seen in Fig. \ref{fig:exp:ula:eigenvaluedist} and \ref{fig:exp:upa:eigenvaluedist}. Even more, the ULA and UPA arrangements are disposed to give different harden supports. To visualize these effects, Fig. \ref{fig:chnhard} was conceived with a reference curve based on uncorrelated fading.

From Fig. \ref{fig:chnhard:r}, it is important to bear in mind that $r$ affects negatively the channel hardening effect as expected because the greater the $r$ is, the stronger the spatial correlation is. One can still note that the UPA provides a minor harden effect than ULA, as also conjectured. In average, the UPA has a variance gain of $1.32$ and $1.93$ times greater than ULA for a moderate, $r=0.5$, and a relative strong, $r=0.75$, correlation circumstances. The upper bound of $r = 1.0$ gives a constant variance equals to $1$, denoting that channel hardening is unattainable at the strongest spatiality case for both array configurations. Although both arrangements have the same upper bound, it has to be said that the UPA strives to achieve this superior limit quicker than the ULA case. Another way to visualize the spatial correlation impact over channel hardening is to set a variance of $10^{-2}$ as a desirable value to be reached. This boundary is considered suitable to support the benefits from the channel hardening effect \cite{Bjornson2017}. Under a moderate spatial correlation scenario of $r=0.5$, the UPA requires $M \approx 256$ to attain this bound, while $M \approx 170$ is required for ULA and $M = 100$ for uncorrelated fading.

The large-scale fading variations over the array are also harmful for the harden level as noted by Fig. \ref{fig:chnhard:std}, being evident that the growth of $\sigma$ is detrimental for the sustenance of channel hardening. As before, it was observed that UPA has a smaller level of channel hardening; seeing that the variance values for UPA are in average $1.32$ and $1.34$ times greater than the ULA equivalent for $\sigma = 4$ dB and $\sigma = 6$ dB, respectively. Assuming the moderate case for $\sigma$ of $4$ dB, the following $M$s are needed to achieve a variance of $10^{-2}$: $M \approx 380$, $M \approx 225$ and $M = 100$ for UPA, ULA, and uncorrelated fading, successively. Thus, if a moderate scenario for the double-parameter model is considered, i.e., $r = 0.5$ and $\sigma = 4$ dB, UPA will require $M \approx 668$ to reach the given bound, whereas $M \approx 296$ will be necessary to ULA and $M = 100$ under uncorrelated fading. 

One can conclude that the fusion of exponential correlation model with large-scale fading variations over the array has a severe impact on the channel hardening effect regarding the uncorrelated case. Besides, the UPA channels always reach a worse performance of channel hardening than the provided by ULA. Notice that the former array needs, in average, $2.25$ more antennas than ULA to reach a variance level of $10^{-2}$; above all, it requires a $6.68$-fold in $M$ to attain the performance of the uncorrelated case. Remember that these given values are assuming a moderate spatial correlation condition.

\section{Impact of Spatial Correlation on the Channel Estimation and Pilot Contamination}\label{sec:spatialcorr-chnest}
The goal of this section is to evaluate how the exponential correlation with large-scale fading variations over the array can influence the channel estimation and pilot contamination of a desired UE. These assessments will be carried out through the realization of illustrative results under particular conditions that attempt to isolate the effect under evaluation.

\subsection{Impact of Spatial Correlation over Channel Estimation}\label{sec:spatialcorr-channel}
\begin{figure}[htp]
	\centering
	\subfloat[NMSE for MMSE estimator as a function of $r$ when $\sigma=0$ dB.]{%
		\includegraphics[trim={6mm 2mm 6mm 5mm}, clip, width=.75\textwidth]{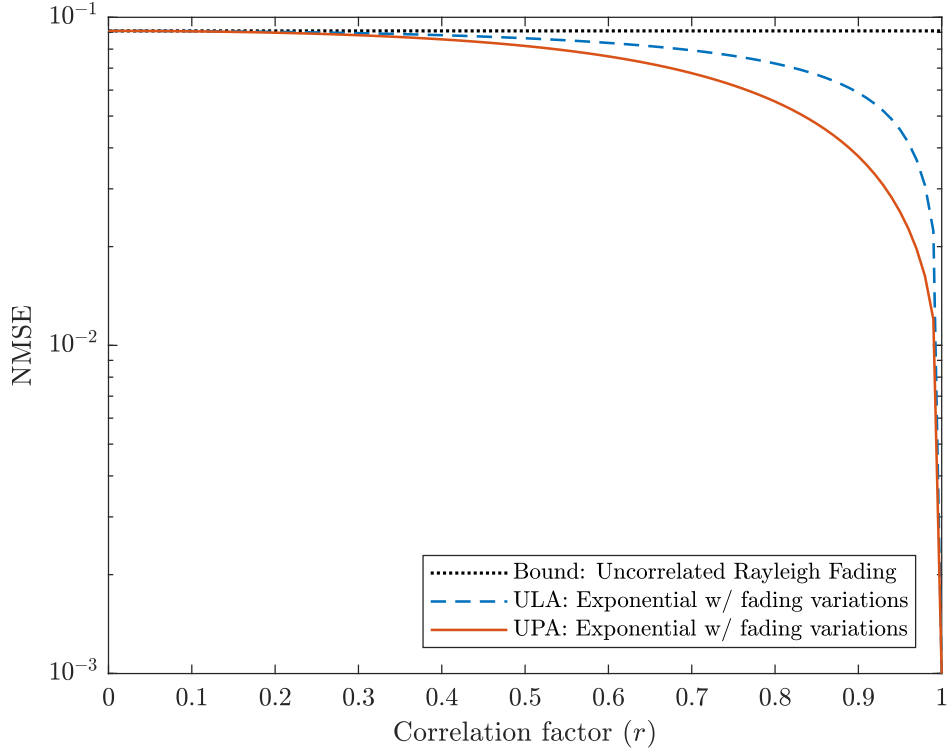} \label{fig:chnest:nmse-spatial:r}}
	
	\subfloat[NMSE for MMSE estimator as a function of $\sigma$ when $r=0$.]{%
		\includegraphics[trim={6mm 2mm 6mm 5mm}, clip, width=.75\textwidth]{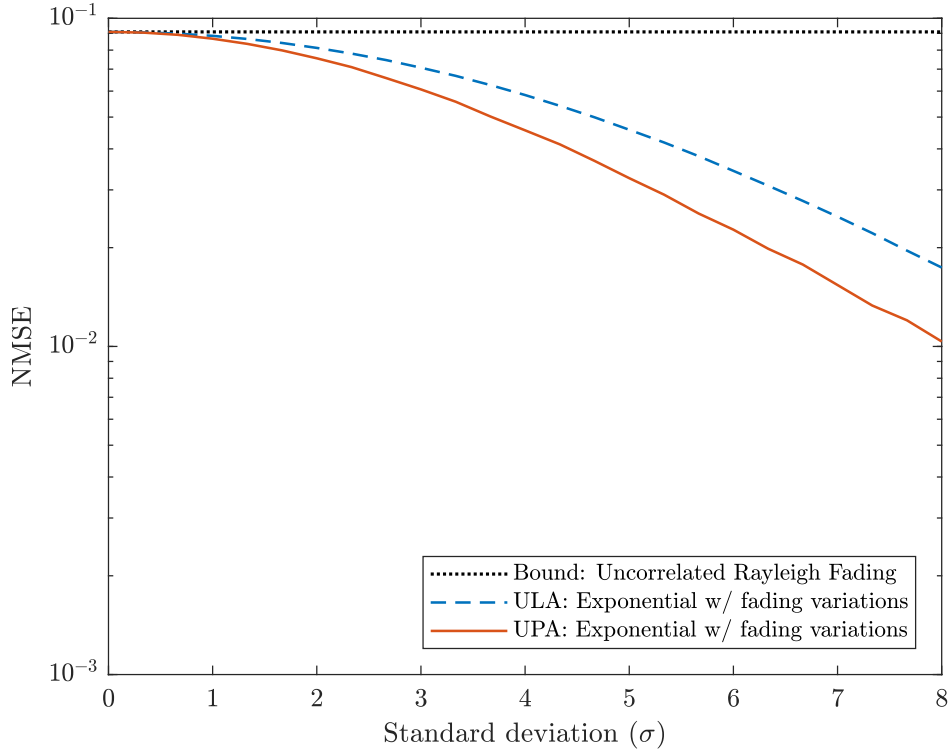}%
		\label{fig:chnest:nmse-spatial:std}
	}
	\caption{NMSE for {MMSE channel estimation} of a desired UE in a free-pilot contamination scenario {when considering} $M=100$ antennas {($M_h=M_{v}=10$)} and an effective SNR $= 10$ dB, as a function of: (a) $r$ when $\sigma=0$ dB and (b) $\sigma$ when $r=0$. The curves were averaged over the nominal angles $\theta$ and $\varphi$.}
	\label{fig:chnest:nmse-spatial}
\end{figure}

Throughout this section, it is considered a singular scenario in which only the desired UE is activated. In this way, the channel estimation of this UE will be investigated under the effect of spatial correlation. Consequently, note that the communication of the desired UE is merely noise-limited without the presence of any other interference. An M-MIMO scheme with $M = 100$ antennas is first assumed to be examined, where the quality of channel estimation, evaluated using the NMSE metric stated in \eqref{eq:mmse:nmse}, is obtained for several values of $r$ and $\sigma$. These results are shown in Fig. \ref{fig:chnest:nmse-spatial} and they are discussed as follows.

Fig. \ref{fig:chnest:nmse-spatial:r} exhibits the NMSE computed for several values of $r$ under both ULA and UPA arrangements. A performance baseline denoted by the uncorrelated Rayleigh fading model was plotted for reference. Moreover, it has to be highlighted that the desired UE has an effective\footnote{The term "effective" SNR stands for the consideration {of the power contribution provided by $\taup$} over the estimation phase.} signal-to-noise-ratio (SNR) of 10 dB. One can observe that the spatiality is not too strong to provide any gain on the channel estimation process until it achieves a correlation degree of approximately $0.4$. As the correlation factor increases towards the fully correlated scenario, $r=1$, a channel estimation error with two orders of magnitude smaller than the uncorrelated case is possible to be obtained through ULA and UPA. Attempting to demonstrate {the effect of the large-scale fading variations over the array} on the channel estimation, Fig. \ref{fig:chnest:nmse-spatial:std} was conceived by holding the same consideration of a desired UE with a SNR $= 10$ dB. Notice that a greater shadowing variability, $\sigma$, generates a stronger spatial correlation level that facilitates the channel estimation procedure. {In summary, the spatial correlation provided by the dual-parameter model is favorable for the improvement of the channel estimation process. This enhancement is fairly greater for the UPA case, owing to the fact that this array has a more confined space between the antenna elements.}

\begin{figure}[htb]
	\centering
	\includegraphics[trim={5mm 2mm 12mm 5mm}, clip, width=.75\textwidth]{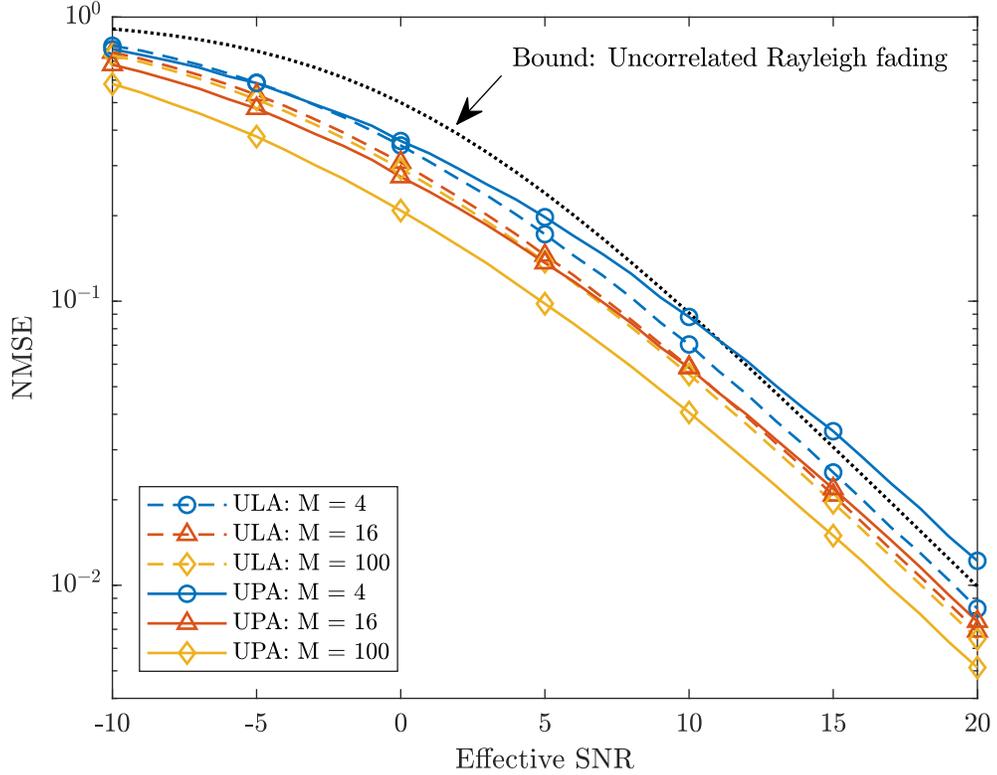}
	\caption{NMSE for MMSE channel estimation as a function of the effective SNR for several values of $M$. The desired UE is under a free-pilot contamination scenario. The exponential model with large-scale fading variations under consideration has $r=0.5$ and $\sigma=4$ dB. The UPA was considered on the shape of a square.}
	\label{fig:chnest:nmse-spatial-mvarying-effectivesinr}
\end{figure}

Due to its structure, the spatial correlation aids to improve the {variance of} the estimate in conjunction with the number of antennas \cite{Bjornson2017}. To visualize this effect, Fig. \ref{fig:chnest:nmse-spatial-mvarying-effectivesinr} illustrates the NMSE for different values of $M$ considering the discussed spatial correlation model with $r=0.5$ and $\sigma=4$ dB. It can be verified that as the number of antennas increases, inclining to {an M-MIMO} operation domain, the NMSE for ULA and UPA tends to shrink in relation to the i.i.d. Rayleigh fading. More precisely, it can be said that {the NMSE under uncorrelated Rayleigh channels} can be reduced by an average-factor of $1.59$ times for ULA and $2.14$ times for UPA when considering $M=100$ antennas and spatially correlated channels. {Overall,} it has to be stressed that the NMSE is a monotonically decreasing function with the SNR, and in asymptotic regime, the NMSE approaches zero \cite{Bjornson2017}.

\subsection{Impact of Spatial Correlation over Pilot Contamination}\label{sec:spatialcorr-pilotcont}
Since pilot contamination stems from the case that UEs are using the same pilot sequence, the scenario of a desired UE being contaminated by an interfering UE will be employed throughout this section. Moreover, the analysis is split into two parts underlying the main effects of pilot contamination over channel estimation: a) the channels turn out to be correlated as deduced in \eqref{eq:mmse:ant-avgcorrcoeff} and b) the estimation quality is degraded as seen from \eqref{eq:mmse:estimate}.

\begin{figure}[htp]
	\centering
	\subfloat[$\sigma = 0$ dB.]{%
		\includegraphics[trim={6mm 2mm 6mm 5mm}, clip, width=.75\textwidth]{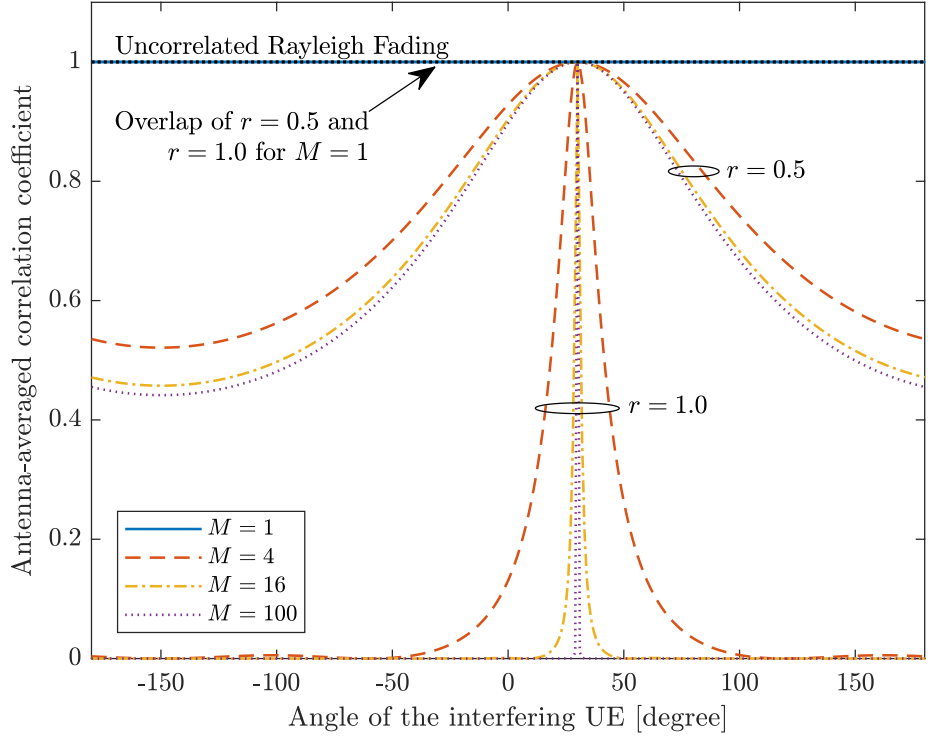}%
		\label{fig:pc:ula:corrcoeff-rvarying}
	}
	
	\subfloat[$r=0.5$.]{%
		\includegraphics[trim={6mm 2mm 6mm 5mm}, clip, width=.75\textwidth]{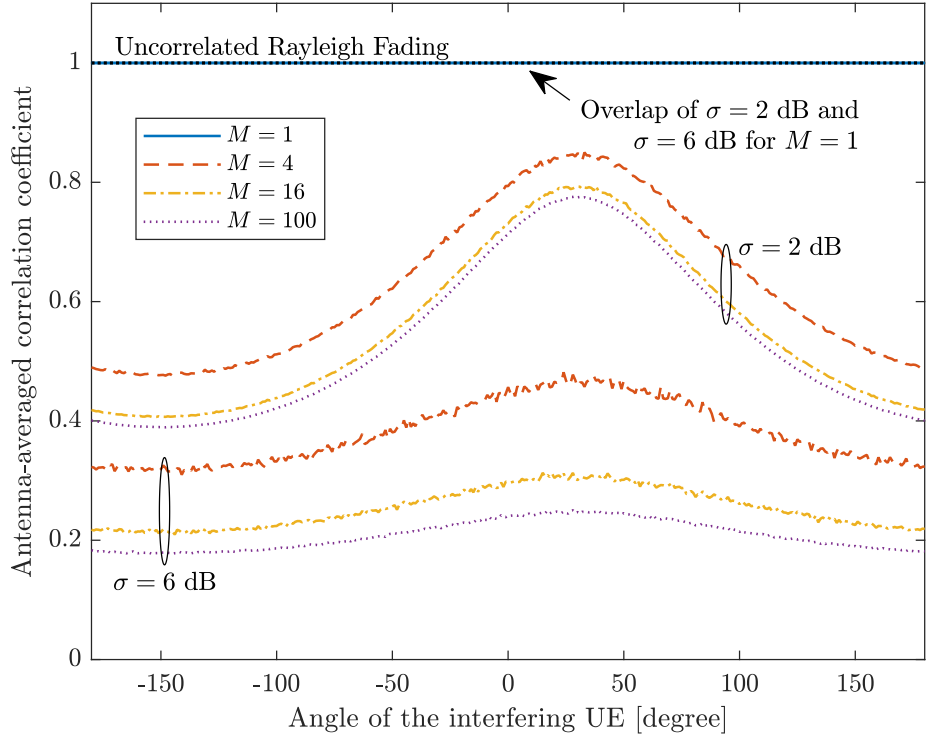}%
		\label{fig:pc:ula:corrcoeff-stdvaying}
	}
	
	\caption{{Antenna-averaged correlation coefficient} as a function of the angular position of the interfering UE for (a) variant $r$ and $\sigma$ fixed and (b) variant $\sigma$ and $r$ fixed. The interfering UE is assigned with the same pilot of the desired UE and it has an effective SNR of $0$ dB. The desired UE has an effective SNR of $10$ dB and a fixed angle of $\theta = 30^{\mathrm{o}}$.}
	\label{fig:pc:ula:corrcoeff}
\end{figure}

The antenna-averaged correlation coefficient is shown in Fig. \ref{fig:pc:ula:corrcoeff-rvarying} for a desired UE fixed $\theta = \pi/3$ with an effective SNR $ = 10$ dB and an interfering UE with variable location over $\theta \in [-\pi,\pi)$ and an effective SNR $ = 0$ dB. This correlation metric is exhibited for the ULA adopting $\sigma = 0$ dB. The figure shows that the larger $M$ is, the smaller the correlation degree between the UEs is, which can be explained by the eigenvalue dependence on $M$ as discussed in Fig. \ref{fig:chnest:nmse-spatial-mvarying-effectivesinr}. Moreover, it can be seen that whenever the interfering UE has the same angle of the desired UE, the correlation coefficient is equal to $1$, i.e., the estimation is fully correlated meaning that the estimates are parallel and they can only be differentiated by a scaling factor. This always occurs for cases of uncorrelated Rayleigh fading and single-antenna ($M=1$). Then, it is crucial to contemplate that, if the UEs have different angles, the correlation level of contamination between UEs is reduced for the correlated model. This highlights the fact that quite different eigendirections of the UEs are beneficial to shrink the correlation between the channel estimates. Another interesting observation is related to the antenna correlation coefficient $r$; when $r$ increases, more spatiality is obtained, and so the more different the channels become, diminishing the contamination between them. Eventually, it can still be seen from Fig. \ref{fig:pc:ula:corrcoeff-rvarying} that the exponential correlation model does not give support to angle resonance at $150^\mathrm{o}$ (reflection angle), which was expected to occur (as seen in \cite{Bjornson2017} for the one-ring model). This effect can be explained resorting on the parameter $r$, by which the desired and interfering $\mathbf{R}$s are made sufficiently distinct in the reflection angle.

Intending to observe the effect of {the large-scale fading variations} over the array, Fig. \ref{fig:pc:ula:corrcoeff-stdvaying} was plotted with $r$ fixed in $0.5$ and variable values of $M$ and $\sigma$. By comparing Fig. \ref{fig:pc:ula:corrcoeff-rvarying} with Fig. \ref{fig:pc:ula:corrcoeff-stdvaying}, it can be noted that the shadow-fading variability causes a distortion in the curves. {In fact, the shadow-fading variations can reduce in average the correlation degree, even if the UEs have similar angles, since these fluctuations strive to differentiate notably the eigenstructure of the UEs due to their randomness.} This phenomenon is expected to be stronger for the UPA, in view of the fact that the shadow-fading has a higher variance under this case, as deduced in Section \ref{sec:exponential}.

\begin{figure}[htb]
	\centering
	\includegraphics[width=0.75\linewidth]{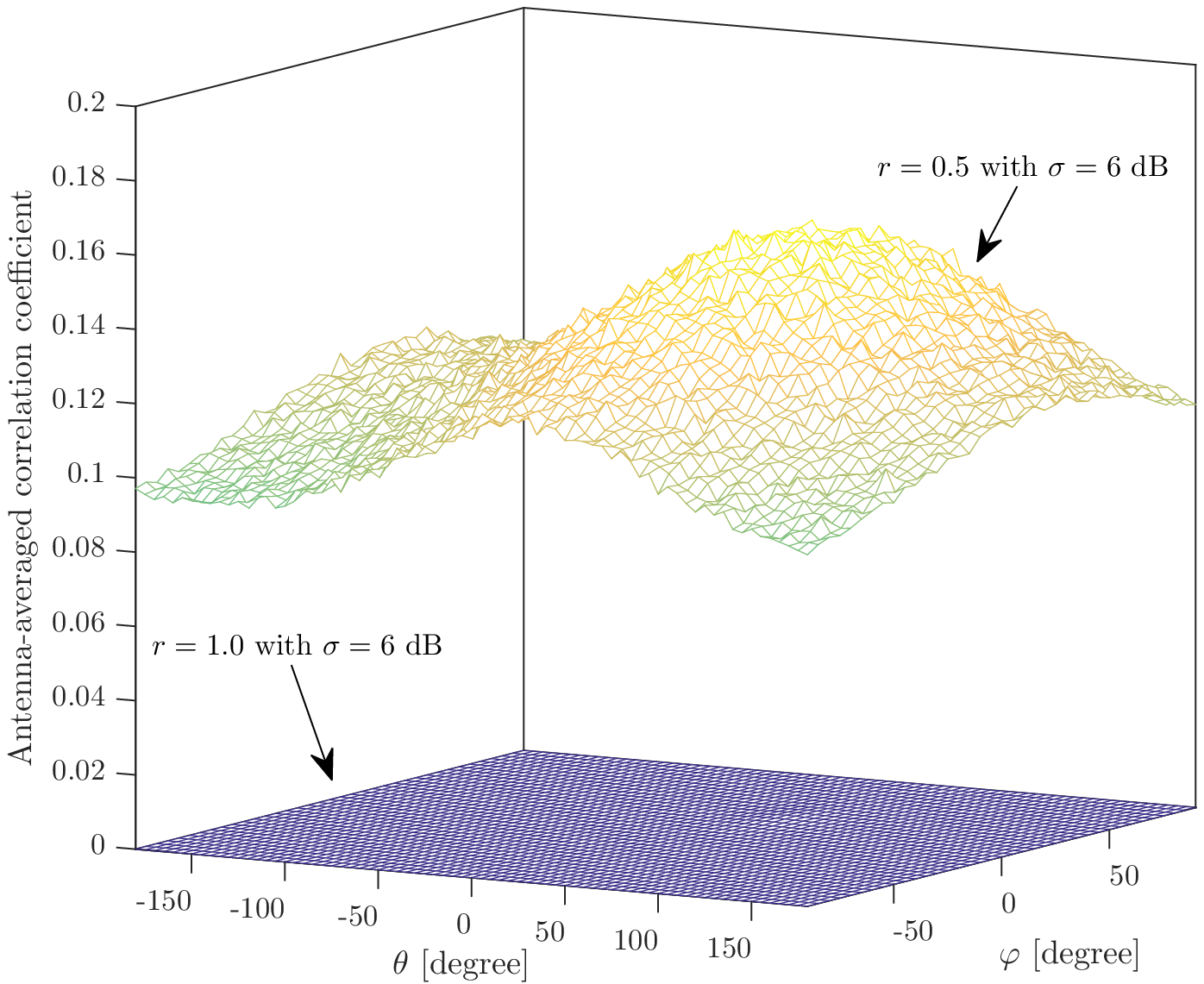}
	\caption{{Antenna-averaged correlation coefficient} as a function of the angular position of the interfering UE for $r = 0.5$ and $r = 1.0$ with $\sigma = 6$ dB. The desired UE has nominal angles equals to $(\theta,\varphi) = (30^{\mathrm{o}},30^{\mathrm{o}})$ and the angles $\theta$ and $\varphi$ of the interfering UE are varying according to $\theta \in [-180^{\mathrm{o}},180^{\mathrm{o}})$ and $\varphi \in [-90^{\mathrm{o}},90^{\mathrm{o}})$. The UPA is equipped with $M = 100$ and the effective SNR of the interfering UE is $0$ dB, whereas the effective SNR of the desired UE is $10$ dB.}
	\label{fig:pc:corrcoeff:upa}
\end{figure}

Fig. \ref{fig:pc:corrcoeff:upa} shows the antenna-averaged correlation coefficient for UPA considering a desired UE at $(\theta,\varphi) = (30^{\mathrm{o}},30^{\mathrm{o}})$ {and an interfering UE varying its position through} $\theta \in [-180^{\mathrm{o}},180^{\mathrm{o}})$ and $\varphi \in [-90^{\mathrm{o}},90^{\mathrm{o}})$. For ease of exposition, the figure only considers a square UPA with $M=100$ antennas. Note that the contamination level is tightly reduced for the UPA case with regard to the equivalent analysis conducted in Fig. \ref{fig:pc:ula:corrcoeff} for the ULA. Eventually, the contamination between the UEs is quasi-nonexistent for $r=1.0$. It is therefore possible to say that the UPA exhibits a reduction of the pilot contamination effect as a result of its superior spatiality, which by its turn reduces the correlation level between the desired and interfering UEs. Furthermore, one can claim that the worst-case scenarios of equal or quasi-equal angles are more difficult to reach when using UPA. This is evidently supported by its degree of freedom being more numerous since, now, the UEs are differentiated by $\theta$ and $\varphi$. To demonstrated the effectiveness of the UPA in mitigating pilot contamination, the mean correlation floor is approximately $0.21$ and $0.12$ for ULA and UPA under $r = 0.5$ with $\sigma = 6$ dB.

Now, we will evaluate the channel estimation quality through the NMSE metric and consider a fixed scenario of $r = 0.5$ and $\sigma = 4$ dB for $M = 100$ antennas when using both ULA and UPA arrangements. The motivation behind these values is the suburban deployment case reported in \cite{Bjornson2018}, which presents a moderate level of spatial correlation. This context also supports some of the spatiality benefits seen above, as can be inferred from the curves given in the next figures of this section.

Fig. \ref{fig:pc:nmse-ula} depicts the NMSE metric as a function of the angle of the interfering UE for the ULA. Several scenarios with different strength of contamination are being evaluated by changing the effective SNR of the interfering UE. The correlated fading cases always provide better channel estimates, owing to the reduction of pilot contamination afforded by spatial correlation. Thus, sufficiently distinct covariance matrices neglect the pilot contamination effect in some order, which fact aids {the obtainment} of a better channel estimation performance. As the interfering power becomes weaker, its impact on the NMSE and the contamination level of the desired UE becomes lesser.

\begin{figure}[htp]
	\centering
	\includegraphics[trim={5mm 2mm 13mm 5mm},clip,width=0.75\linewidth]{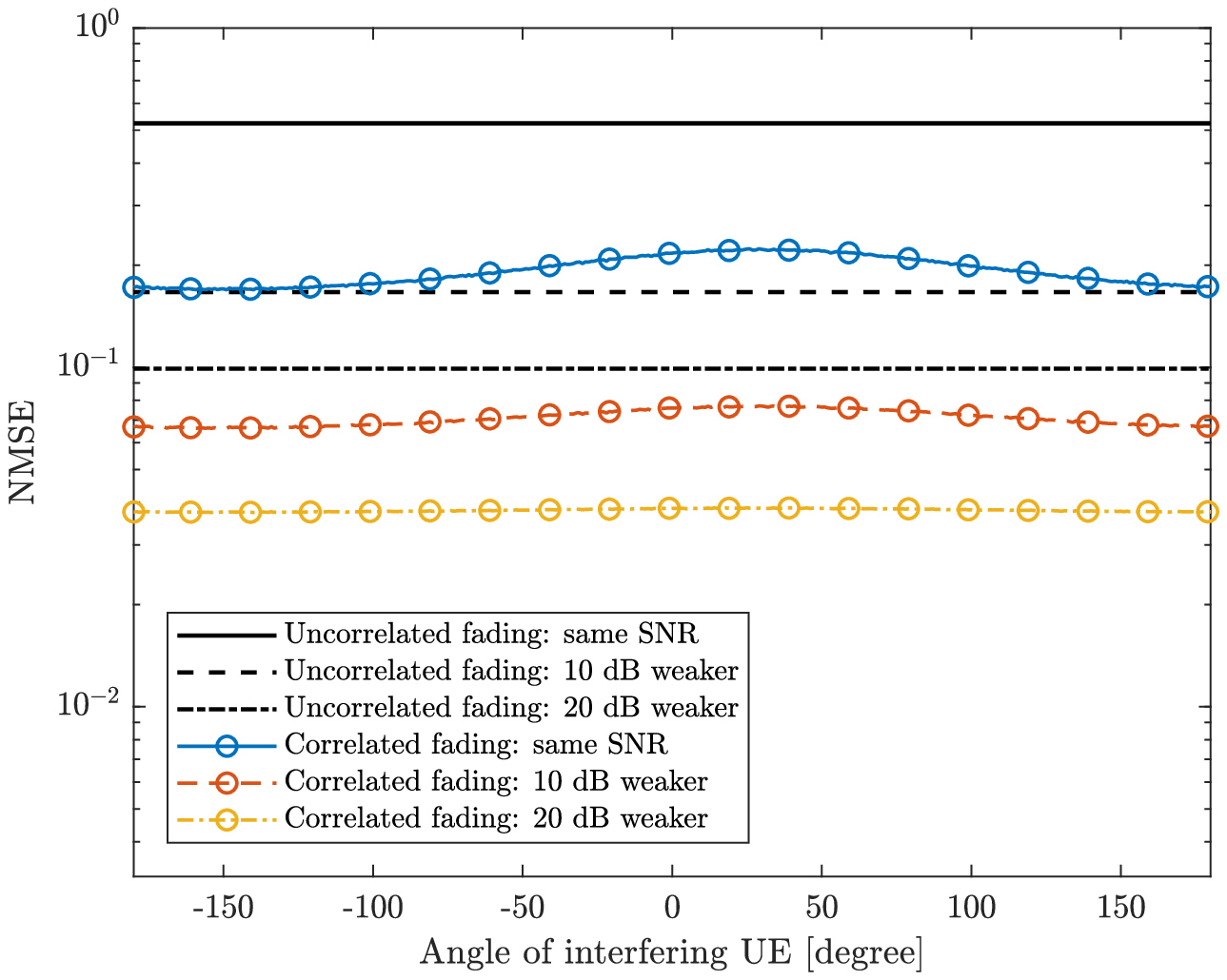}
	\caption{NMSE of the desired UE as a function of the azimuth angle of the interfering UE. The UE of interest is located at $\theta = 30^{\mathrm{o}}$ with effective SNR of $10$ dB, whereas the interfering UE has an azimuthal angular variation between $-180^{\mathrm{o}}$ to $180^{\mathrm{o}}$. The ULA is equipped with $M = 100$ antennas under moderate spatial correlation of $r=0.5$ and $\sigma=4$ dB. Three levels of power difference between the desired UE and the interfering UE were considered.}
	\label{fig:pc:nmse-ula}
\end{figure}

\begin{table}[htp]
	\centering
	\caption{Summary of the results obtained for ULA and UPA arrangements regarding channel hardening and channel estimation (favorable propagation) metrics under uncorrelated and correlated fading channels.}
	\label{tab:summaryresults}
	\begin{tabular}{|c|c|c|c|c|}
		\hline
		\textbf{} & \textbf{\begin{tabular}[c]{@{}c@{}}Channel \\ Hardening\end{tabular}} & \multicolumn{3}{c|}{\textbf{\begin{tabular}[c]{@{}c@{}}Channel Estimation\\ (Average NMSE)\end{tabular}}} \\ \hline
		\textbf{Scenario} & \textit{\begin{tabular}[c]{@{}c@{}}$M$ needed to reach \\ a variance of $10^{-2}$\end{tabular}} & \textit{\begin{tabular}[c]{@{}c@{}}Same \\ SNR\end{tabular}} & \textit{\begin{tabular}[c]{@{}c@{}}10 dB \\ weaker\end{tabular}} & \textit{\begin{tabular}[c]{@{}c@{}}20 dB\\ weaker\end{tabular}} \\ \hline
		\textit{\begin{tabular}[c]{@{}c@{}}Uncorrelated \\ Rayleigh Fading\end{tabular}} & $100$ & $0.5238$ & $0.1667$ & $0.0991$ \\ \hline
		\textit{\begin{tabular}[c]{@{}c@{}}ULA with moderate \\ spatial correlation\end{tabular}} & $296$ & $0.1930$ & $0.0710$ & $0.0379$ \\ \hline
		\textit{\begin{tabular}[c]{@{}c@{}}UPA with moderate\\ spatial correlation\end{tabular}} & $668$ & $0.0667$ & $0.0305$ & $0.0189$ \\ \hline
	\end{tabular}
\end{table}

\begin{figure}[htb]
	\centering
	\includegraphics[width=0.75\linewidth]{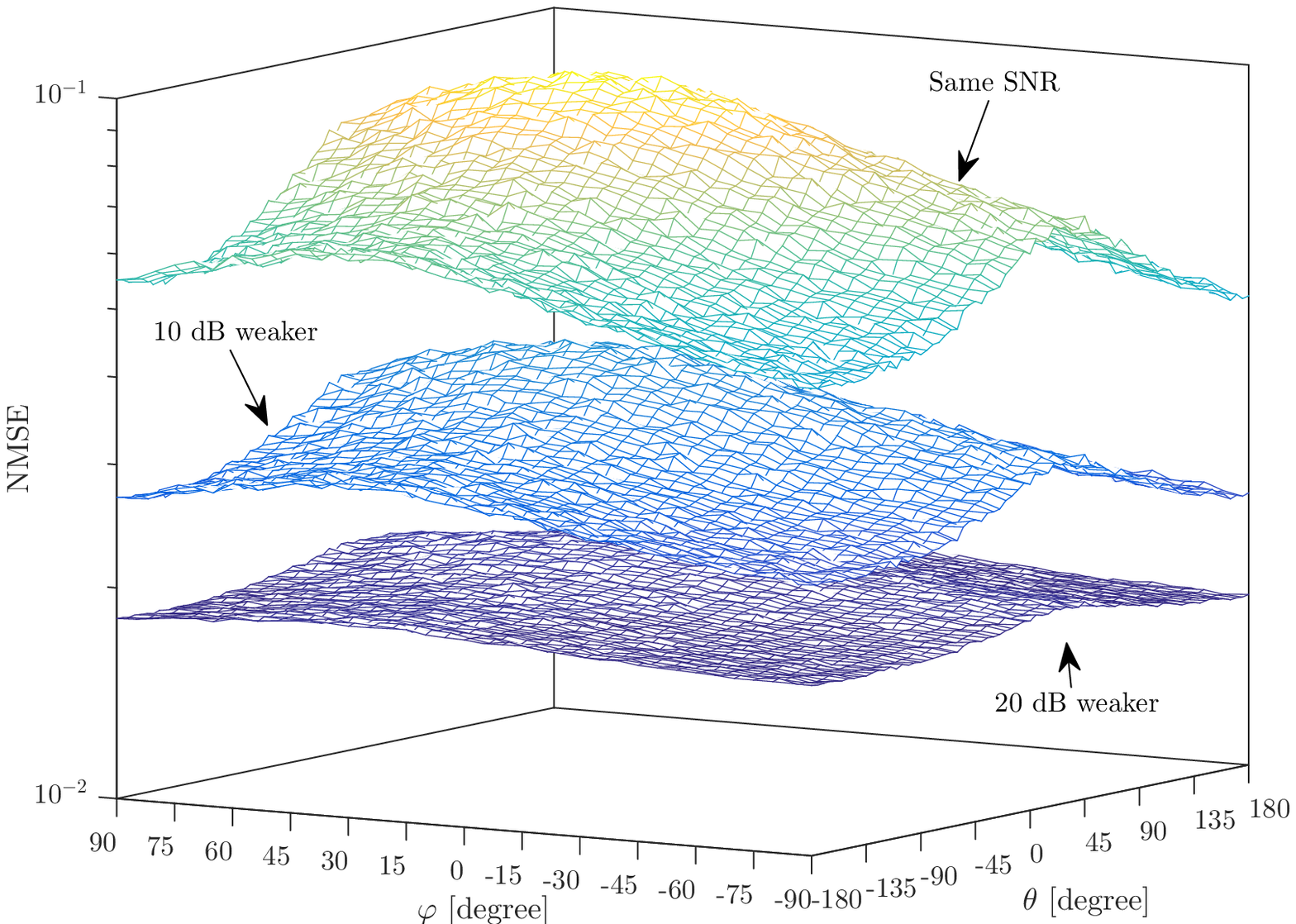}
	\caption{NMSE of the desired UE as a function of the azimuth and the elevation angles of the interfering UE. The UE of interest is located at $(\theta,\varphi) = (30^{\mathrm{o}},30^{\mathrm{o}})$ with effective SNR of $10$ dB, whereas the interfering UE is varying its position following $\theta \in [-180^{\mathrm{o}},180^{\mathrm{o}})$ and $\varphi \in [-90^{\mathrm{o}},90^{\mathrm{o}})$. The square UPA is equipped with $M = 100$ antennas under moderate spatial correlation of $r=0.5$ and $\sigma=4$ dB. Three levels of power difference between the desired UE and the interfering UE were considered.}
	\label{fig:pc:nmse-upa}
\end{figure}

To observe the estimation quality offered by the UPA, in Fig. \ref{fig:pc:nmse-upa}, the NMSE of the desired UE is plotted as a function of the azimuth and the elevation angles of the interfering UE. Like in Fig. \ref{fig:pc:nmse-ula}, there are several scenarios denoting different strengths of contamination. Besides, the uncorrelated Rayleigh fading was omitted to facilitate the graphical disposition. As expected, the UPA presents better NMSE results than the ULA arrangement, which is demonstrated in Fig. \ref{fig:pc:nmse-ula}. More precisely, one can gather the following NMSE average-gain values for UPA in relation to ULA: $2.89$, $2.33$, and $2.01$, for the cases of same SNR, $10$ dB weaker, and $20$ dB weaker, respectively. The UPA can also reach NMSE values almost one order of magnitude lower than the uncorrelated case in all the scenarios. 

Table \ref{tab:summaryresults} summarizes the main results of the whole paper regarding the channel hardening and favorable propagation effects, where the latter was indirectly assessed by the NMSE metric of the channel estimation. This table lists the number of antennas required to reach the $10^{-2}$ channel hardening bound of the variance stated in \eqref{eq:variance:chnhardening} when considering the uncorrelated and moderately correlated fading scenarios. Moreover, it shows the average NMSE values acquired for the three different cases of interfering power strength, once again under none or moderate level of spatial correlation with $r=0.5$ and $\sigma=4$ dB. Note that, as the values of the average NMSE decrease for ULA and UPA, the value of $M$ needed to attain the channel hardening bound increases. For this reason, one can clearly see a trade-off between the assurance of the channel hardening effect and the enhancement of the channel estimation process. It turns out that, as large as the antenna array becomes, its implementation also comes to be more expensive and spacious. In summary, the choice of the antenna array type has to take into account, mainly, the cost and space constraints for the ULA and UPA deployment cases, and then verify which arrangement is the most beneficial under a given circumstance.

\section{Conclusions}\label{sec:conc}
In this paper, a compound of the exponential correlation model with large-scale fading variations over the array was duly assessed for ULA and UPA arrangements. This double-parameter model consists of an interesting way to design the spatial correlation in some practical circumstances of M-MIMO channels. Then, the channel hardening and favorable propagation effects were analyzed numerically, indicating that both are influenced by the assumed spatial correlation model. Moreover, it was possible to observe the mitigation of pilot contamination when the values of $r$ and $\sigma$ are combined. This mitigation level was seen potentially stronger for the UPA, where an average gain of almost $2$-fold better NMSE values were crudely\footnote{The term "crudely" is used because the given average gain disregards the interfering power.} encountered regarding the ULA case and under moderate spatial correlation.

Although spatiality is beneficial to the estimation quality, it contributes poorly to the assurance of the channel hardening effect, where a counter-gain of an almost $2$-fold was observed for UPA w.r.t. the ULA equivalent; again, under a moderate spatial correlation condition. Thus, the most crucial design point observed was the trade-off between favorable propagation and channel hardening, which should be handled with care when deciding which antenna arrangement to use in an M-MIMO implementation case.

%\newpage

%%%%%%%%%%%%%%%%%%%%%%%%%%%%%%%%%%%%%%%%%%%%%%%%%%%%%%%%%%%%%%%%%%%%%%%%%%%

\vspace{2mm}
%%%%%%%%%%%%%%%%%%%%%%%%%%%%%%%%%%%%%%%%%%%%%%%%%%%%%%%%%%%%%%%%%%%%%%%%%%%%%%%%
%	References
%%%%%%%%%%%%%%%%%%%%%%%%%%%%%%%%%%%%%%%%%%%%%%%%%%%%%%%%%%%%%%%%%%%%%%%%%%%%%%%%
%\bibliographystyle{IEEEtran}
%\bibliography{refs/exp_corr}


%%%%%%%%%%%%%%%%%%%%%%%%%%%%%%%%%%%%%%%%%%%%%%%%%%%%%%%%%%%%%%%%%%%%%%%%%%%%

\end{document}